\documentclass[prb,twocolumn,preprintnumbers,amsmath,amssymb]{revtex4}
\usepackage{graphicx}% Include figure files
\usepackage{epsfig}
\usepackage{dcolumn}% Align table columns on decimal point
\usepackage{bm}% bold math
\usepackage{times}
\usepackage{mathptmx}
\usepackage{color}
\usepackage{amsmath}
\usepackage{xfrac}
\usepackage{amsfonts}
\usepackage{xspace}

\begin{document}

\title{Photoexcited states of the harmonic honeycomb iridate $\gamma-$Li$_2$IrO$_3$}

\author{J. P. Hinton}
\affiliation{Materials Science Division, Lawrence Berkeley National Laboratory, Berkeley, California 94720, USA}
\affiliation{Department of Physics, University of California, Berkeley, California 94720, USA}
\affiliation{Department of Physics, University of California, San Diego, California 92093, USA}

\author{S. Patankar}
\affiliation{Materials Science Division, Lawrence Berkeley National Laboratory, Berkeley, California 94720, USA}
\affiliation{Department of Physics, University of California, Berkeley, California 94720, USA}

\author{E. Thewalt}
\affiliation{Materials Science Division, Lawrence Berkeley National Laboratory, Berkeley, California 94720, USA}
\affiliation{Department of Physics, University of California, Berkeley, California 94720, USA}

\author{J. D. Koralek}
\affiliation{Materials Science Division, Lawrence Berkeley National Laboratory, Berkeley, California 94720, USA}

\author{A. Ruiz}
\affiliation{Materials Science Division, Lawrence Berkeley National Laboratory, Berkeley, California 94720, USA}
\affiliation{Department of Physics, University of California, Berkeley, California 94720, USA}

\author{G. Lopez}
\affiliation{Materials Science Division, Lawrence Berkeley National Laboratory, Berkeley, California 94720, USA}
\affiliation{Department of Physics, University of California, Berkeley, California 94720, USA}

\author{N. Breznay}
\affiliation{Materials Science Division, Lawrence Berkeley National Laboratory, Berkeley, California 94720, USA}
\affiliation{Department of Physics, University of California, Berkeley, California 94720, USA}

\author{I. Kimchi}
\affiliation{Department of Physics, University of California, Berkeley, California 94720, USA}

\author{A. Vishwanath}
\affiliation{Materials Science Division, Lawrence Berkeley National Laboratory, Berkeley, California 94720, USA}
\affiliation{Department of Physics, University of California, Berkeley, California 94720, USA}

\author{J. Analytis}
\affiliation{Materials Science Division, Lawrence Berkeley National Laboratory, Berkeley, California 94720, USA}
\affiliation{Department of Physics, University of California, Berkeley, California 94720, USA}

\author{J. Orenstein}
\affiliation{Materials Science Division, Lawrence Berkeley National Laboratory, Berkeley, California 94720, USA}
\affiliation{Department of Physics, University of California, Berkeley, California 94720, USA}

\begin{abstract}
We report equilibrium and nonequilibrium optical measurements on the recently synthesized "harmonic" honeycomb iridate $\gamma-$Li$_2$IrO$_3$ (LIO), as well as the layered honeycomb iridate Na$_2$IrO$_3$ (NIO). Using Fourier transform infrared microscopy we performed reflectance measurements on LIO, from which we obtained the optical conductivity below 2 eV. In addition we measured the photoinduced changed in reflectance, $\Delta R$, as a function of time, $t$, temperature, $T$, and probe field polarization in both LIO and NIO. In LIO, $\Delta R(t,T)$ is anisotropic and comprised of three $T$ dependent components. Two of these components are related to the onset of magnetic order and the third is related to a photoinduced population of metastable electronic excited states. In NIO, $\Delta R(t,T)$ has a single $T$ dependent component that is strikingly similar to the electronic excitation component of $\Delta R$ in LIO. Through analysis and comparison of $\Delta R(t,T)$ for two compounds, we extract information on the onset of magnetic correlations at and above the transition temperature in LIO, the bare spin-flip scattering rate in equilibrium, the lifetime of low-lying quasiparticle excitations, and the polarization dependence of optical transitions that are sensitive to magnetic order.
\end{abstract}

\date{\today}

\maketitle

\section{Introduction}

Transition metal oxides (TMO) host complex phases resulting from interactions with a hierarchy of energy scales. In $3d$ and $4d$ TMOs, competition between kinetic energy, quantified by the parameter, $t$, and Coulomb repulsion, parameterized by $U$, are the major factors determining the nature of lowest energy phases.  However, in $5d$ systems the spin-orbit (SO) interaction, which plays a subsidiary role in the lighter TM’s, becomes an equal partner in shaping the nature of the electronic states.

The iridate family of TMO's are a particularly striking example of interplay between SO and $U$ interactions. It is proposed that strong SO interaction reorganizes the crystal field states of the $5d$ orbitals into a $J$-multiplet structure, where $J$ is the combined spin and orbital angular momentum. The relatively weak $U$ is then sufficient to produce localization in the singly occupied $J=1/2$ doublet, giving rise to a novel Mott insulator in which the local moments have both spin and orbital character.~\cite{KimPRL08,KimScience09} Based on an effective single-band Hubbard model obtained by projection onto the $J=1/2$ subspace, it was predicted that the layered perovskite iridates are analogs of cuprate parent compounds and could exhibit high-$T_c$ superconductivity when doped.~\cite{SenthilPRL} Recently, evidence for a metallic state with a pseudogap,~\cite{KimScience14} as well as hints of high-$T_c$ superconductivity~\cite{Jan15,Kim15} have been reported in Sr$_2$IrO$_4$ with an overlayer of K, heightening interest in the electronic properties of iridates. However, despite the appeal of the $J=1/2$ picture it remains somewhat controversial, as quantum chemical considerations suggest that SO, $U$, $t$, and crystal field interactions are of comparable magnitude, such that neither a local $J$-multiplet nor delocalized orbital picture is entirely appropriate.~\cite{AritaPRL12,CominPRL12,HaskelPRL12,HseihPRB12,MazinPRL12,MazinPRB13,SohnPRB13}

In addition to the questions concerning the origin of the insulating ground state and low-lying electronic excitations, there is considerable interest in the magnetic correlations in iridates.  In compounds of the form A$_2$IrO$_3$, where A is Na or Li, the combination of strong SO coupling and edge sharing IrO$_6$ octahedra is thought to give rise to anisotropic Kitaev magnetic exchange.~\cite{ChaloupkaPRL10,SinghPRL12}  Na$_2$IrO$_3$ (NIO), which possesses a layered honeycomb structure, was the first iridate to be scrutinized in the search for a realization of the Kitaev spin liquid. Neutron and X-ray diffraction studies revealed instead a rather simple form of magnetic order: a coplanar antiferromagnet with a zigzag structure.~\cite{YePRB12,LiuPRB11,ChoiPRL12} Although recent diffuse X-ray scattering measurements provide direct evidence for the existence of bond-directional magnetic interactions in NIO,~\cite{ChunNatPhys15} this conventional form of magnetic order suggests that Kitaev interactions do not dominate the other symmetry allowed spin couplings in this system.~\cite{FoyevtsovaPRB13,MazinPRL12,KimSciRep14}

\begin{figure}[ht]
	\begin{center}
		\includegraphics[width=8cm]{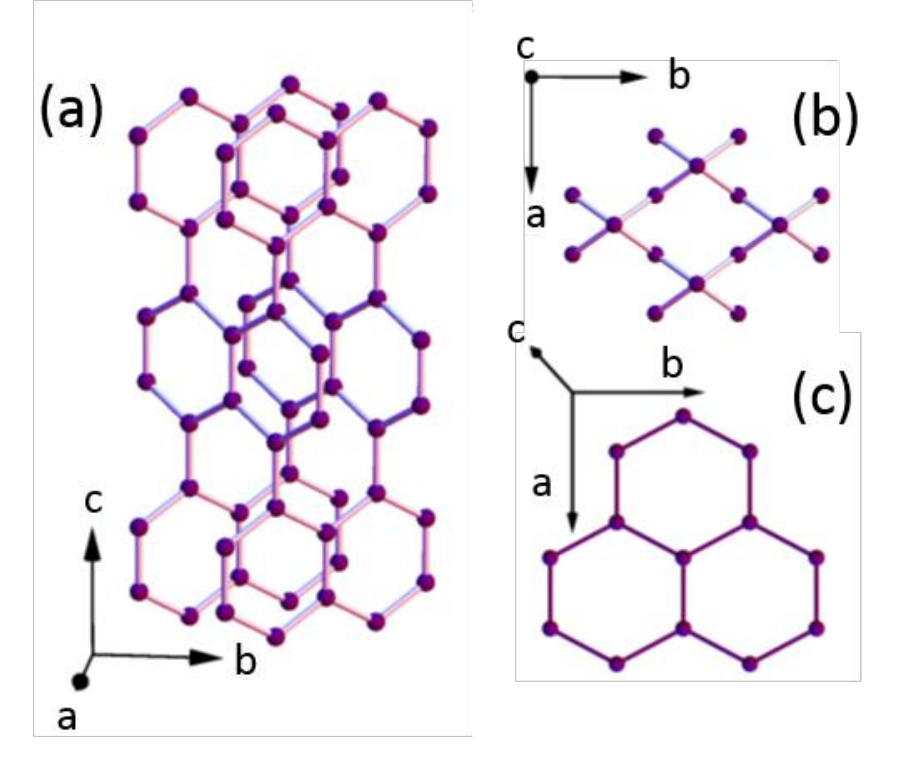}
	\end{center}
	\caption{\label{fig:1} (a) Three dimensional view of the crystal structure of LIO, with only the iridium atoms shown for simplicity. (b) and (c) show the positions of Ir atoms as viewed in the plane from which the reflectance is measured, for LIO and NIO, respectively.}
\end{figure}

Recently, two polytypes of Li$_2$IrO$_3$, $\beta$~\cite{TakayamaArxiv14} and $\gamma$,~\cite{ModicNatCom14} have been synthesized with structures that were previously unknown.  Each of these has the same basic building block of three-fold coordinated Ir ions as the layered honeycomb structure.  However, in both new polytypes the plane formed by the triad of Ir links rotates to create three dimensional rather than layered structures. In contrast to the comparatively conventional magnetic structure in NIO, these “harmonic honeycomb” iridates host pairs of incommensurate, non-coplanar, and counter-propagating spin spirals. ~\cite{BiffinPRL14} Comparison with the ground state of model spin Hamiltonians suggests that Kitaev interactions must dominate Heisenberg terms in order to produce the complex spin spirals that are observed.~\cite{KimchiArxiv14,KimchiPRB14,LeeArxiv14,ReutherArxiv14}

The detailed knowledge of the magnetic order of the new iridate compounds is in sharp contrast with our understanding of their electronic structure, both from an experimental and theoretical point of view.  In the case of the layered honeycomb iridates, the results of RIXS, PES, and optical spectroscopy, considered together with $T$-dependence of the resistivity, point to an insulating state with a bandgap on the order of $0.5$~eV. ~\cite{SinghPRB10,CominPRL12,GretarssonPRL13,SohnPRB13} Far less is known about $\gamma-$Li$_2$IrO$_3$ (LIO) because of its recent discovery, as well as the relatively small ($100~\mu$m) dimensions of crystals synthesized to date.

In the work reported here we probed electronic excitations using two optical methods. We used Fourier transform infrared microscopy to perform broadband polarized measurements of the infrared reflectivity on LIO single crystals, from which we extract the two in-plane components of the optical conductivity tensor in the photon energy, $\hbar\omega$, range from $0.2$ to $2.0$ eV. In addition, we have performed transient optical reflectivity measurements on both LIO and NIO, probe the dynamics of magnetic ordering and photoexcited quasiparticles.

\section{Equilibrium optical conductivity}

\begin{figure}[ht]
	\begin{center}
		\includegraphics[width=8cm]{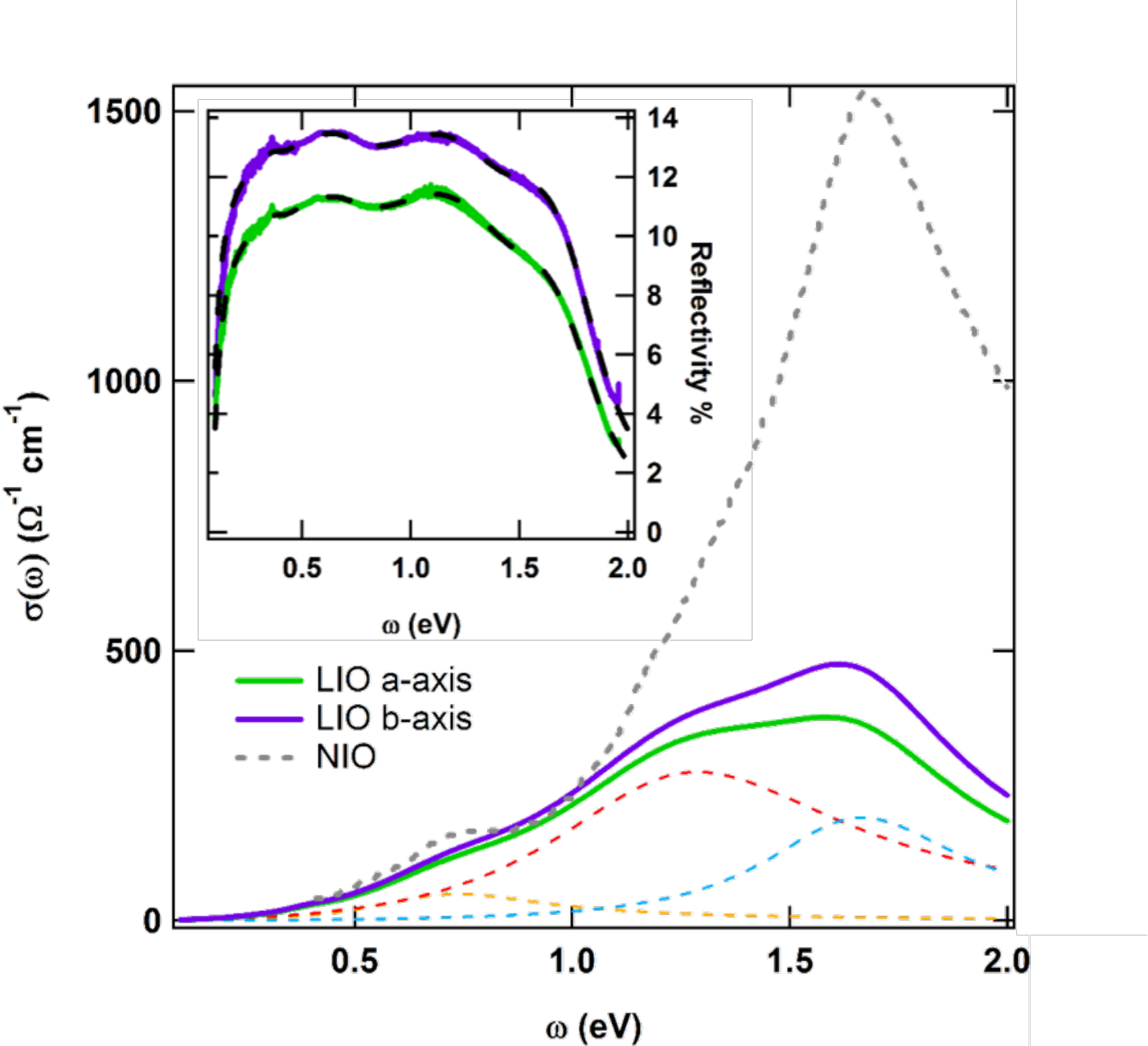}
	\end{center}
	\caption{\label{fig:2} Optical conductivity of LIO and NIO. The green and purple curves correspond to the \textit{a} and \textit{b}-axis optical conductivity in LIO. Optical conductivity of NIO from Reference~\cite{SohnPRB13} is shown by the gray dashed line. The colored dashed lines show the contributions from Lorentz oscillators used to fit the LIO reflectance curves shown in the inset.}
\end{figure}

Figs. 1a and 1b illustrate the atomic structure of LIO. Fig. 1a provides a 3D perspective on the overall crystal structure, showing only the Ir atoms for clarity.  The structure is seen to consist two sets of chains of hexagons, oriented in the directions \textbf{a}$\pm$\textbf{b}.  A hexagon that lies in one set of such chains is connected to its nearest neighbor chain in the other set through an Ir-Ir link oriented in the \textit{c}-direction.  The surface of the crystals suitable for optical microscopy is the plane perpendicular to \textbf{c}, as shown in Fig. 1b. The simple 2D honeycomb structure of NIO is illustrated in Fig. 1c.

\begin{table}
	\caption{\label{tab:table1} Fit parameters for the optical reflectivity data in Fig. 2, with all values given in cm$^{-1}$. The background dielectric constants used in the fit are $\epsilon^{a}_{\infty} = 2.24$ and $\epsilon^{b}_{\infty} = 2.56$.}
	\begin{ruledtabular}
		\begin{tabular}{ccccc}
			& $\omega_{0}$ & $\Gamma$ & $\omega^{a}_{p}$ & $\omega^{b}_{p}$\\
			\hline
			Phonon & 248 & 0.997 & 991 & 985\\
			\hline
			Interband & 3154 & 983 & 515 & 493\\
			& 5968 & 3829 & 3224 & 3483\\
			& 10540 & 6929 & 10691 & 11119\\
			& 13506 & 4310 & 6935 & 8374\\
		\end{tabular}
	\end{ruledtabular}
\end{table}

The spectra in Fig. 2 show that the anistropy in the equilibrium reflectivity, $R(\omega)$, is rather small.  The fits to $R(\omega)$ were obtained by modeling the optical conductivity, $\sigma(\omega)$, as a sum of contributions from four Lorentz oscillators at $0.4$, $0.7$, $1.3$, and $1.7$ eV.  Excellent fits were obtained for the two principal axes using the same resonance frequency and damping but slightly different oscillator strengths, with fit parameters given in Table 1. The $\sigma(\omega)$ thus obtained (shown as solid lines) is quite similar to that found at low energy in NIO ($<1$ eV), but is substantially smaller at higher energy. Calculations of $\sigma(\omega)$ have been performed within density functional theory (including both SO and $U$ couplings) for the layered honeycomb polytype of Li$_2$IrO$_3$ and Na$_2$IrO$_3$.~\cite{LiArxiv14} The spectra that emerge from this theory are quite similar for these two compounds, suggesting that the large difference we observe is a consequence of the inherently 3D structure of $\gamma-$Li$_2$IrO$_3$, rather than the replacement of Na by Li.

\section{Transient reflectance}

\subsection{$\Delta R(t,T)$ in LIO an NIO}

\begin{figure}[ht]
	\begin{center}
		\includegraphics[width=8cm]{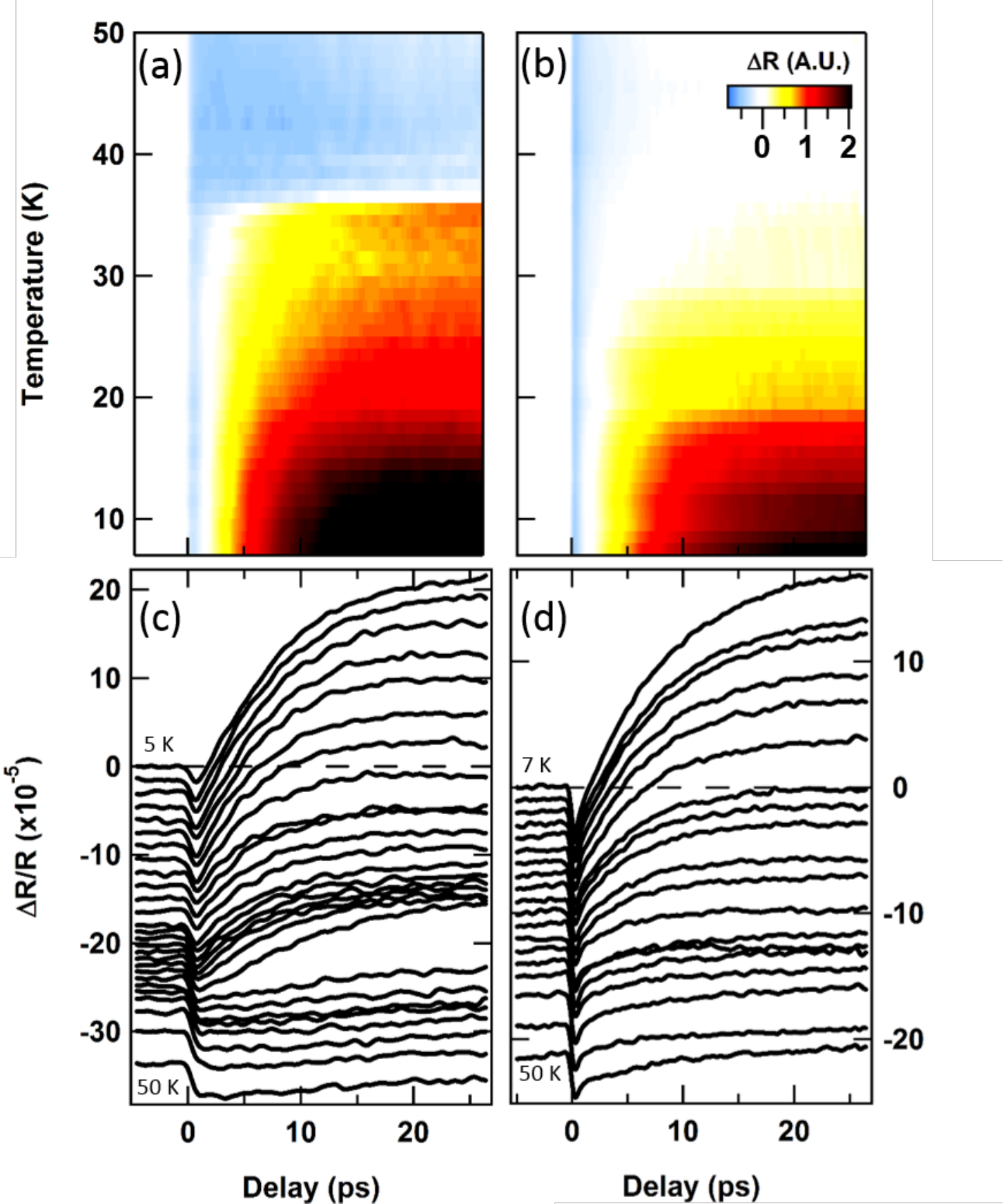}
	\end{center}
	\caption{\label{waterfall} $\Delta R(t,T)$ as a false color image for LIO along the \textit{a}-axis (a) and for NIO (b). The same data is plotted as a series of curves for LIO (c) and NIO (d).}
\end{figure}

Measurements of $\Delta R$ were performed with 100 fs pulses of 800 nm light, with pump and probe beams focused to a 100 micron spot on the sample.  The measurements were performed in a low pump power regime,  1 $\mu$J/cm$^2$, where the magnitude of $\Delta R$ is a linear function of pump fluence and the rise and decay times for $\Delta R$ are independent of fluence.  The dependence of $\Delta R(t,T)$ on temperature, $T$, and delay time, $t$, for LIO and NIO are illustrated in Fig. \ref{waterfall}. $\Delta R$ in the $t-T$ plane is shown as a false color image in Figs. \ref{waterfall}a and \ref{waterfall}b for LIO and NIO, respectively, and in waterfall plots of $\Delta R(t)$ for various temperatures in the range 5-50K (Figs. \ref{waterfall}c and \ref{waterfall}d). For the LIO data, the probe was polarized in the \textbf{a}-direction (data for probe parallel to \textbf{b} will be described below). In NIO, $\Delta R(t,T)$ is independent of probe polarization as expected for electric fields in the \textit{ab}-plane of the layered hexagonal structure.

\begin{figure}[ht]
	\begin{center}
		\includegraphics[width=8 cm]{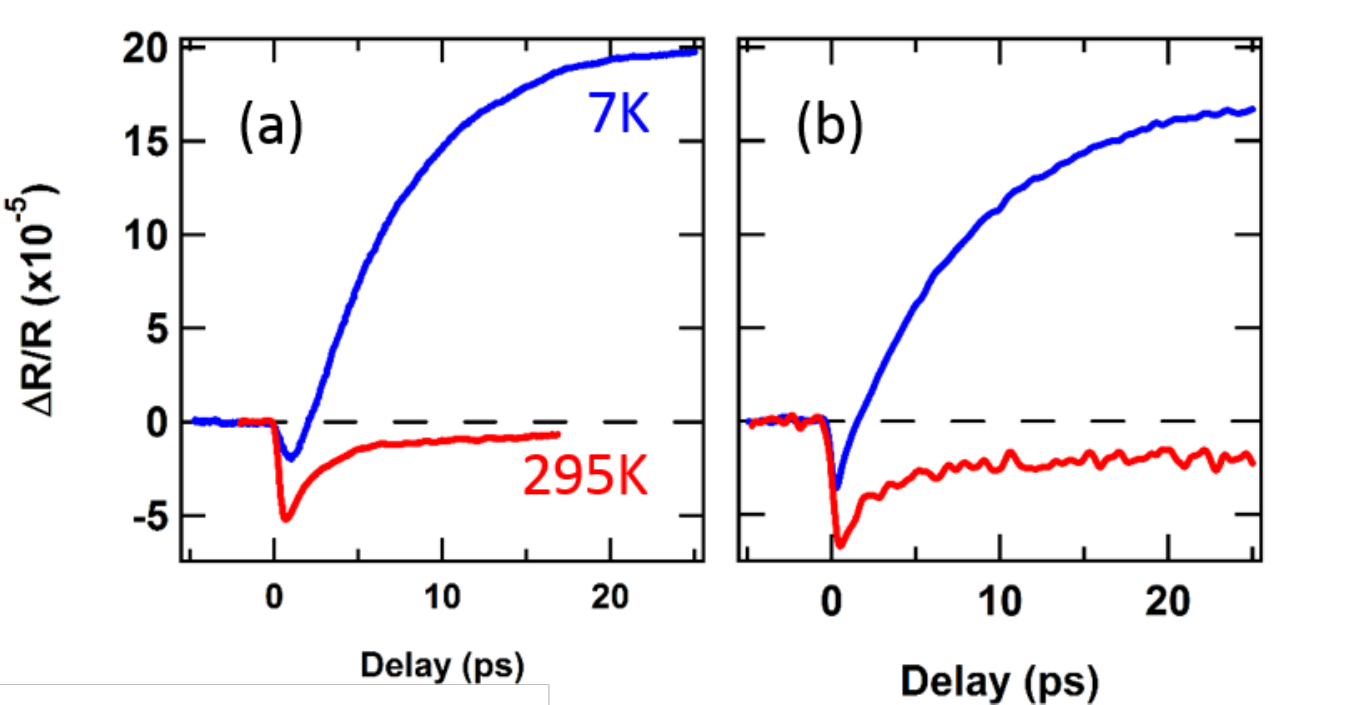}
	\end{center}
	\caption{\label{5K_295K}Pump-probe reflectivity at room temperature (red) and $5$ K (blue) is shown along LIO \textit{a}-axis in (a) and for NIO in (b).}
\end{figure}

To introduce the comparison of transient photoreflectance in the two compounds, we first plot $\Delta R(t)$ at the high and low $T$ limits of our measurement range. Figs. \ref{5K_295K}a and \ref{5K_295K}b show $\Delta R(t)$ of the two compounds at 5 K and 295 K.  The magnitude, sign, and time dependence of $\Delta R$ in both compounds are strikingly similar at these two temperatures. At 295 K, $\Delta R(t)$ is negative with an abrupt onset of less than $1$ ps duration. The initial rapid decay of $\Delta R(t)$ likely reflects the cooling of photoexcited electrons and holes as they reach quasi-thermal equilibrium with the lattice.  The ultimate return to equilibrium of the coupled electron-phonon system then takes place on the nanosecond timescale.  At low $T$, in both NIO and LIO, the negative reflectivity transient is accompanied by a larger positive component of $\Delta R(t)$ that rises on much slower, $\sim 10$ ps, time scale.

\begin{figure}[ht]
	\begin{center}
		\includegraphics[width=8.48cm]{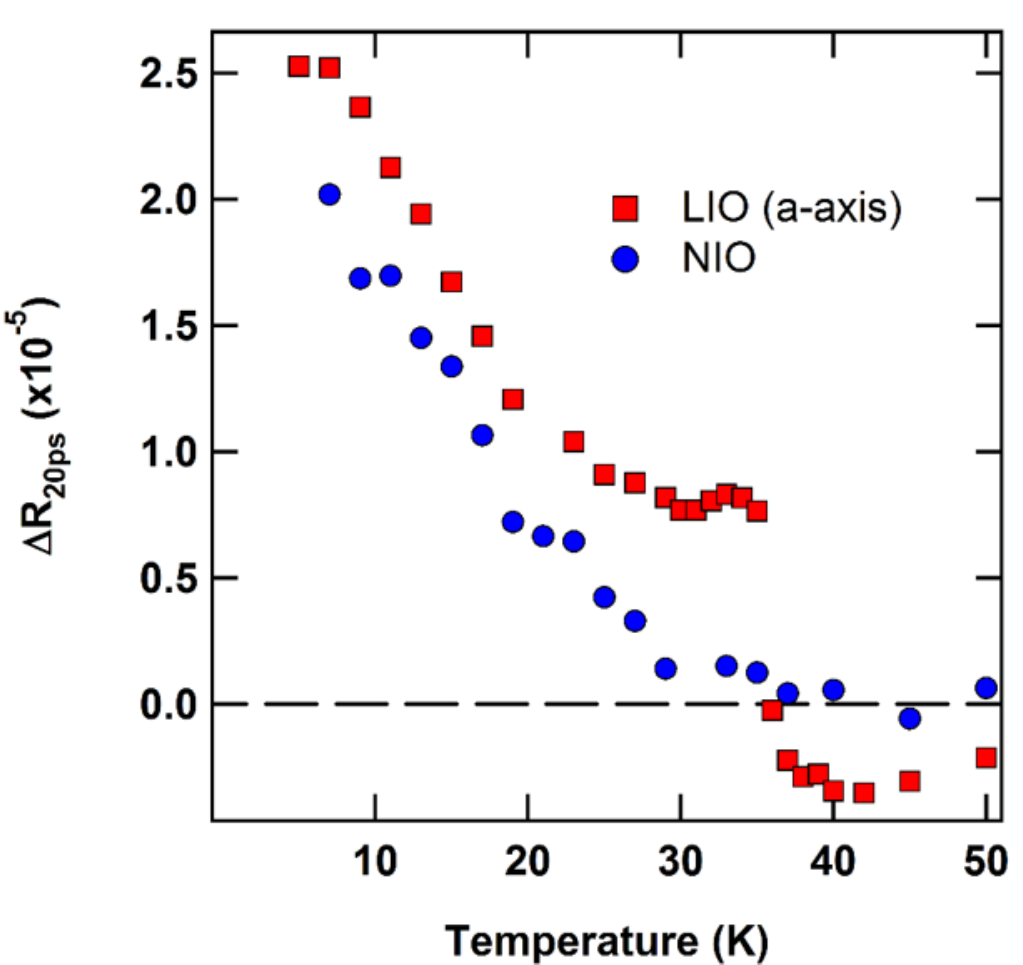}
	\end{center}
	\caption{\label{20ps}$\Delta R(t,T)$, sampled at a fixed time delay of 20 ps, is plotted as a function of temperature. $\Delta R(20 \textrm{ps},T)$  is nearly identical for the two compounds below $30$ K, but note the abrupt jump in the LIO response near its $T_c$ of 36 K, which is absent in NIO at its $T_c$ of 14 K.}
\end{figure}

Although $\Delta R(t)$ is virtually the same in LIO and NIO at 5 K and 295 K, the pathway from low to high temperature is quite different in the range of $T$ where magnetic order appears. While this contrast is already apparent in Fig. \ref{waterfall}, we can illustrate the difference more clearly by comparing $\Delta R(T)$ for the two compounds sampled at a fixed time delay of $t = 20$ ps (Fig. \ref{20ps}). In NIO the positive component of $\Delta R(T)$ emerges gradually upon cooling below $\sim 40$ K whereas in LIO an abrupt change appears at its magnetic transition temperature, $T_c$, of $36$ K.

To describe the time dependence of $\Delta R$ at low $T$ the positive component was isolated by subtracting $\Delta R(t, T = 50$~K$)$ from the curves at lower $T$. The resulting single component signals can be accurately fit with an exponential rise of the form, $\Delta R(t)=A(1-e^{-t / \tau_r})$. In Figure \ref{taurise} we plot $\tau_r (T)$ thus obtained for both compounds.  The contrasting behavior of LIO and NIO seen in $\Delta R(20ps,T)$ appears clearly in $\tau_r(T)$ as well.  Whereas in NIO the rise time increases smoothly with decreasing $T$, in LIO it diverges at $T_c$.

\subsection{Critical component in LIO}

The behavior of $\Delta R(T)$ and $\tau_r(T)$ near $T_c$ in LIO has been observed in many systems that undergo second-order magnetic phase transitions,~\cite{KisePRL00,OgasawaraPRL05,KantnerPRB11} and is seen whenever the order parameter can be probed by reflectance. A phenomenological, mean-field model that accounts for the enhanced amplitude and divergent time scale of $\Delta R$ was developed by Koopmans \textit{et al.}~\cite{KoopmansPRL05} Their analysis is based on the assumption that the pump pulse increases the effective temperature of the spin system, $T_s$. The reduction in spin order $\delta S$ corresponding to an increase in spin temperature $\delta T_s$ approaches $\delta S\propto (T_c-T)^{-1/2}\delta T_s$ as $\delta T_s\rightarrow 0$, which diverges at the transition. In actual experiments the divergence is broadened by both disorder and the nonzero value of $\delta T_s$.

\begin{figure}[ht]
	\begin{center}
		\includegraphics[width=8cm]{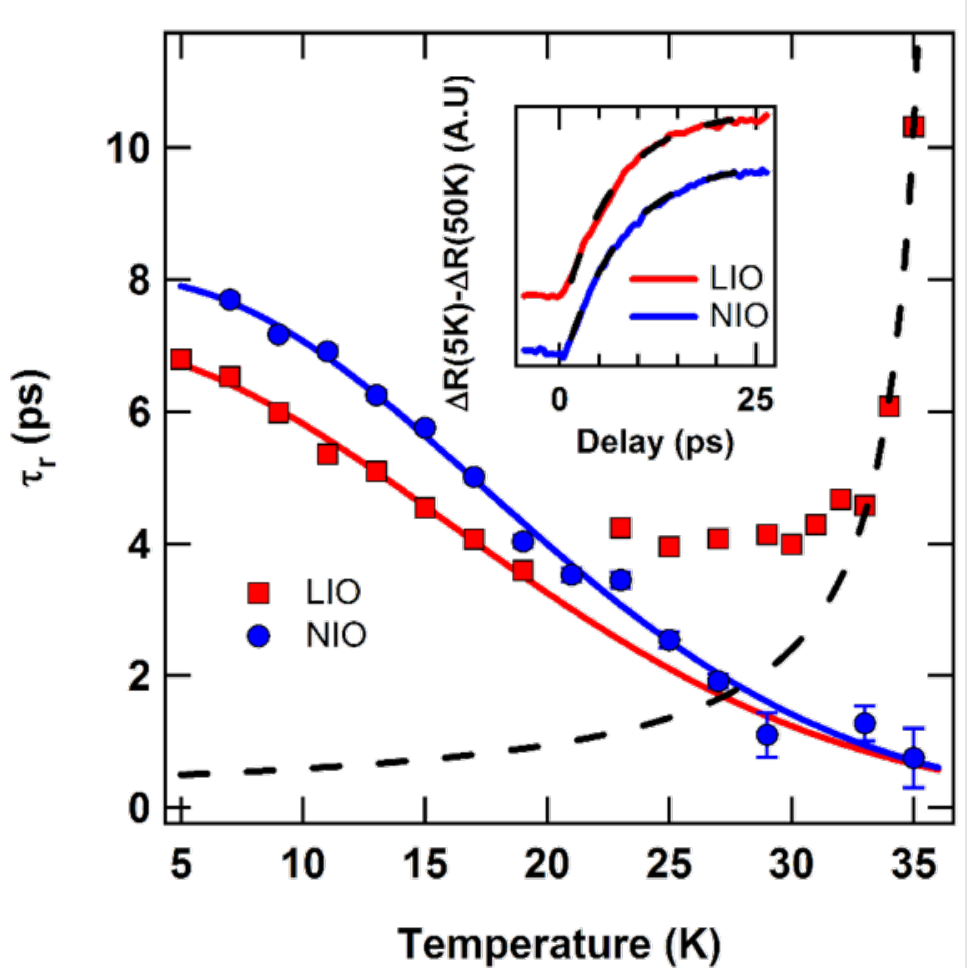}
	\end{center}
	\caption{\label{taurise}Temperature dependence of the exponential rise time of $\Delta R$. The inset shows the slow-rising component of $\Delta R$, isolated by subtracting the data at $T = 50$ K, with fits in black dash. Note that $\tau_r$ diverges at $T_c$  in LIO, while there is no feature at the magnetic transition temperature of 14 K in NIO. The smooth temperature dependence of the non-critical component is illustrated by the solid line guides to the eye. The black dashed line is a fit to the model of critical slowing down described in the text.}
\end{figure}

The principle of detailed balance can be used to calculate the rate at which a system with spin order $S(T_s)$ approaches the quasiequilibrium value $S(T_s+\delta T_s)$. In mean-field theory, the time rate of change of $S$ is,
\begin{equation}
\frac{dS}{dt}=2\gamma_{sf}(S_\downarrow e^{-b/T_s}-S_\uparrow),
\end{equation}
where $\gamma_{sf}$ is a bare spin flip rate, $b(T_s)$ is the effective self-consistent exchange field, and $S_\downarrow$ and $S_\uparrow$ are the fraction of spins aligned and anti-aligned with $b(T_s)$, respectively. If at equilibrium at $T_s$, $dS/dt=0$, then following a step-like increase to $T_s+\delta T_s$, $S$ will decrease at a rate given to leading order in $b(T_s)$ by,

\begin{equation}
\frac{dS}{dt}\approx -\gamma_{sf} \left (\frac{b}{T_s^2}\right)\delta T_s.
\end{equation}
The characteristic time to approach quasiequilibrium is therefore,
\begin{equation}
\tau_S\equiv \frac{\delta S}{dS/dt}=\frac{1}{2 \gamma_{sf}} \frac{d\ln b}{d\ln T},
\end{equation}
which yields $\tau_r=\gamma_{sf}^{-1}T_c/2(T_c-T)$ for any power law singularity of $b(T_s)$. The dashed line in Fig. \ref{taurise} shows a fit to this prediction for $\tau_r(T)$ with parameters $T_c=$36.5 K and $\gamma_{sf}=$ 1.75 THz.

\subsection{Non-critical component in LIO and NIO}

The data and analysis presented thus far suggest a description of $\Delta R$ in terms of two contributions, one of which is common to both NIO and LIO, and the other that is unique to LIO. The common component is one that grows gradually in amplitude and appears above the detection threshold at approximately 30 K, while the component that appears only in LIO has a sharp onset at its $T_c$ of 36 K. Below we analyze the non-critical component that is common to both compounds and we discuss possible explanations for the absence of critical component in NIO in section III.D.

\begin{figure}[ht]
	\begin{center}
		\includegraphics[width=8cm]{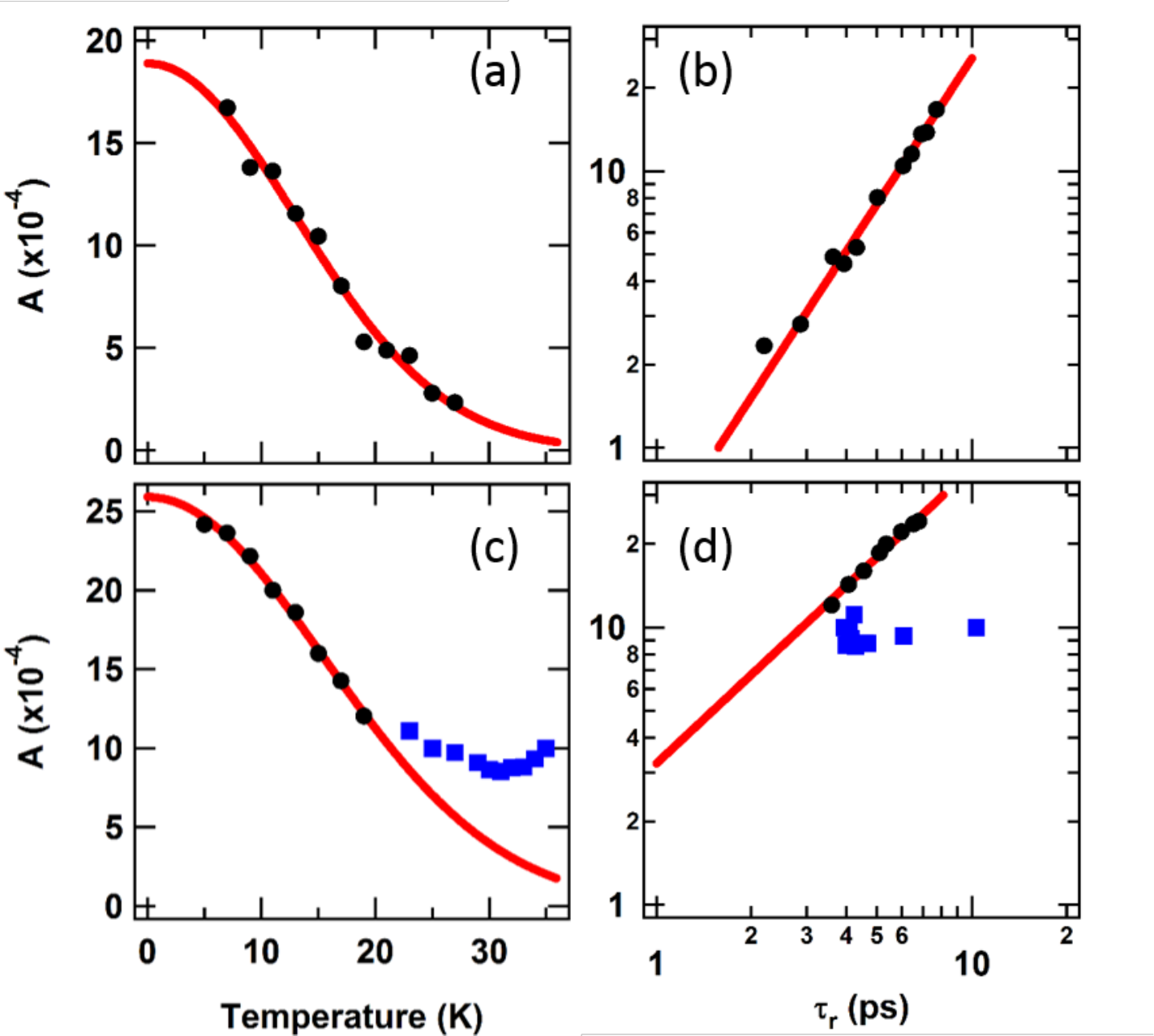}
	\end{center}
	\caption{\label{A_vs_tau}The amplitude $A$ obtained from the fits described in the text is plotted as function of temperature and risetime $\tau_r$ for NIO in (a) and (b) and LIO in (c) and (d). The red lines in (a) and (c) plots are a guide to the eye showing the smooth temperature dependence of $A$. The red lines in (b) and (d) show a power law fit to the data, with exponents of 1.75 for NIO and 1.06 for LIO. The blue squares that deviate from the power law dependence in (d) represent data in the critical regime, $T>$ 20 K.}
\end{figure}

In Figs. \ref{A_vs_tau}a and \ref{A_vs_tau}c we plot $A(T)$ obtained from the fit procedure described above for NIO and LIO, respectively. The solid line is a guide to the eye that has the same shape in both plots and illustrates the commonality and contrast in the NIO and LIO data. In NIO the amplitude and rise time of $\Delta R$ increase monotonically with decreasing $T$.  The correlation between the two parameters is shown in the double logarithmic plot of $A(T)$ vs. $\tau_r(T)$ (Fig. \ref{A_vs_tau}b.), where the straight line fit indicates the relation $A(T)\propto[\tau_r(T)]^{1.75}$. In LIO this correlation is observed only at temperatures below the critical regime, as shown in Fig. \ref{A_vs_tau}d. The black dots correspond to temperatures below 25 K and fall on a straight line of slope 1.06, indicating an approximately linear relationship between $A$ and $\tau_r$. On the other hand, the points shown as blue squares, which correspond to 25~K $<T<$ 40~K clearly have a different relationship between amplitude and risetime that is related to criticality.

The correlation between the amplitude and rise time in the non-critical regime can be described by a simple kinetic model, whose basis is independent of the microscopic nature of the photoexcitations. First, we recognize that excited states are created simultaneously with the absorption of the pump photons.  Therefore, the relatively slow risetime of $\Delta R$ indicates that the initial photoexcited state, which we label 1, does not generate a measurable change in reflectance at the probe photon energy of 1.5 eV. We assume that state 1 can either decay directly to the ground state, or to a metastable state (labelled 2) that generates a $\Delta R$ at 1.5 eV in proportion to its population. Such a kinetic model is described mathematically by two coupled rate equations of the form,

\begin{equation}
\frac{dN_1}{dt}=\Phi(t)-\frac{N_1}{\tau_1}-\gamma_{12}N_1,
\end{equation}
\begin{equation}
\frac{dN_2}{dt}=\gamma_{12}N_1,
\end{equation}
where $N_1$ and $N_2$ are the initial and secondary excited state densities, respectively, $1/\tau_1$ is the rate for direct decay of excitation 1 to the ground state, and $\gamma_{12}$ is the rate of conversion of species 1 to 2. The solution for pumping term $\Phi(t)=F\delta(t)$ is,
\begin{equation}
N_2(t)=F\frac{\gamma_{12}}{\gamma}(1-e^{-\gamma t}),
\end{equation}
where $\gamma\equiv (\tau_1^{-1}+\gamma_{12})$. In the limit that $\tau_1^{-1}\gg\gamma_{12}$, the metastable population of state 2 approaches $F\gamma_{12}\tau_1$ and $\tau_r\rightarrow\tau_1$. Therefore if the conversion rate is independent of temperature, the model predicts a linear relationship between the amplitude and risetime of $\Delta R$.  The physical picture for the correlation between $A(T)$ and $\tau_r$ is that, with decreasing $T$, the increased lifetime of the initial (but invisible) state allows more time for build-up of the secondary (but observable) state. As previously noted, this linear relationship between $A(T)$ and $\tau_r$ holds for LIO, while in NIO we observe a power law relationship with an exponent of 1.75. This difference could arise from $T$-dependence in NIO of $\gamma_{12}$ or the proportionality of $\Delta R$ to $N_{2}$, both of which could potentially give rise to a power law different from one over the relatively small $T$ range from 7 to 27 K.

\subsection{Anisotropy}

In this section we describe the dependence of $\Delta R$ on the polarization of the probe pulse and discuss the relationship between anisotropy and the absence of the critical component in NIO. Fig. 8a illustrates the strong anisotropy that we observe in LIO by comparing $\Delta R_a (t,T)$ and $\Delta R_b (t,T)$, the change in $R$ with probe field $\mathbf{E}$ parallel to \textbf{a} and \textbf{b}, respectively, for selected temperatures from 10 to 65 K. In contrast, $\Delta R$ is isotropic in NIO, as is illustrated in Fig. 8b.

As the dependence of $\Delta R$ in LIO on polarization is rather complicated, we have used shading in Fig 8a to illustrate how the anisotropy develops with decreasing $T$. In addition, in Fig. 8c we compare the temperature dependence of $\Delta R_a$ and $\Delta R_b$ sampled at $t=$ 20 ps. The negative component of $\Delta R$ observed at high $T$ is isotropic.  Upon cooling below 65 K, $\Delta R_b (t,T)$ varies from the isotropic high-$T$ response, becoming more negative, whereas $\Delta R_a (t,T)$ remains roughly $T$ independent. However, as $T$ reaches $T_c$, the nature of the anisotropy changes with the appearance of the large critical component discussed above. Although the step-like change in $\Delta R$ is observed for both directions of $\mathbf{E}$, the amplitude of the step is approximately 2-3 times larger for $\mathbf{E}$ parallel to \textbf{a}, as illustrated in Fig. 8c.

\begin{figure}[ht]
	\begin{center}
		\includegraphics[width=8.5cm]{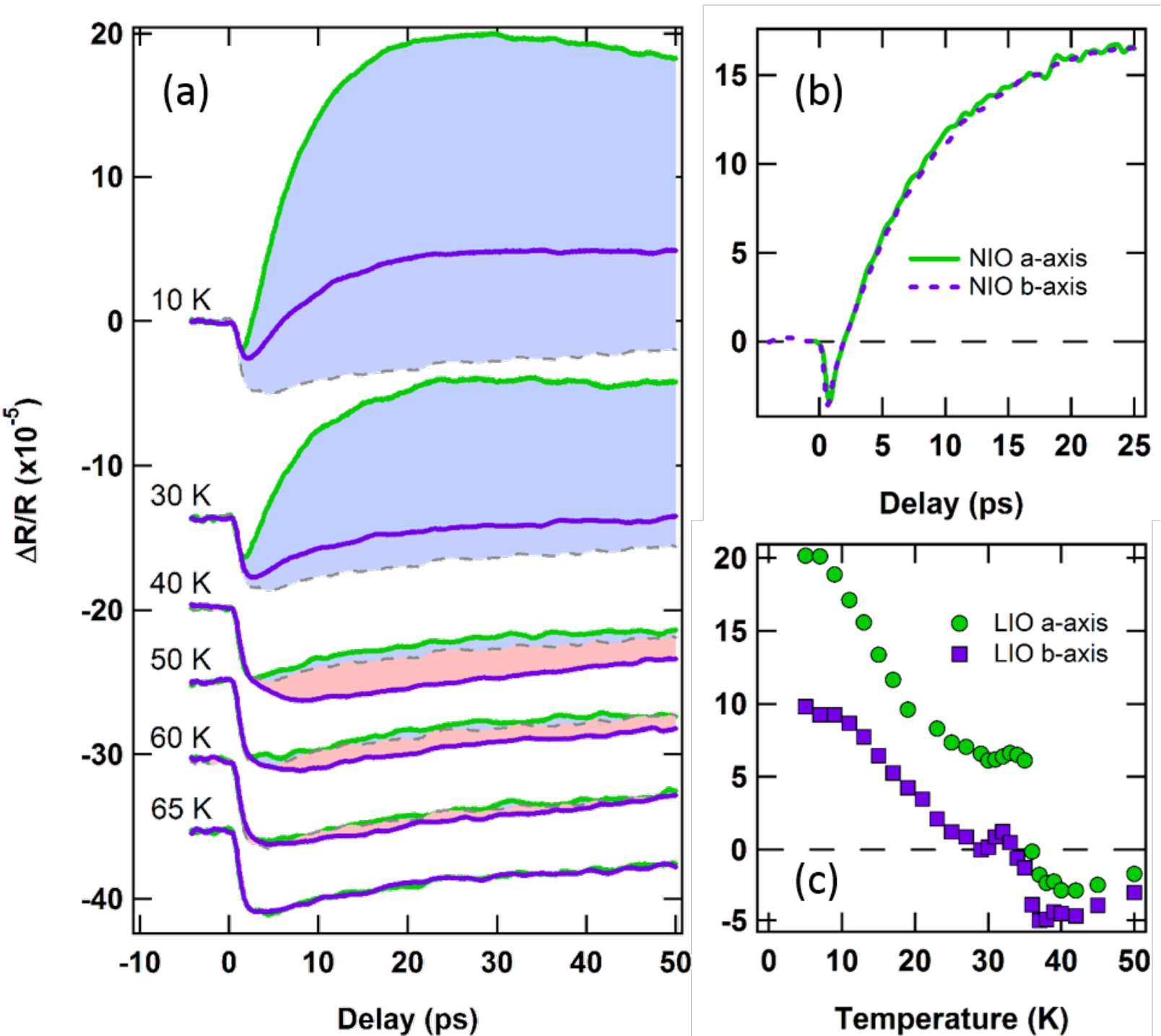}
	\end{center}
	\caption{\label{fig:4} (a) The pump-probe response in LIO with probe-polarization along the \textit{a} and \textit{b} axes is shown for a series of temperatures spanning $T_c$. The \textit{a}-axis data for each temperature is shown in green, with the \textit{b}-axis data in purple. The dashed gray lines show the $70$ K data as reference, with the blue shading indicating a positive difference from the $70$ K signal and red shading indicating a negative difference. (b) The isotropic nature of $\Delta R$ in NIO illustrated by reflectance transients observed at 7 K. $\Delta R(t=20$~ps$,T)$ for the probe polarization parallel to \textbf{a} and \textbf{b} is plotted in (c).}
\end{figure}

The contrast in the polarization dependence of the transient reflectivity in LIO and NIO is very likely related to to the difference in the orientation of the hexagonal Ir units with respect to the crystallographic axes and to the surface from which $R$ is measured. In NIO, the Ir hexagons are coplanar, forming a honeycomb network in the plane containing the probe field. In LIO, the Ir hexagons form two sets of chains.  Looking down on the reflecting \textit{a-b} plane surface, there are two sets of hexagons viewed edge on. As stated earlier, the normal directions  for the two sets are oriented in directions \textbf{a}$\pm$\textbf{b}, respectively. Thus, in the case of LIO, $R$ samples transition dipole moments both parallel and perpendicular to the hexagonal planes, while in NIO only the in-plane polarized transitions are detected. This basic structural difference immediately suggests an explanation for the missing critical component in NIO: \textit{the component of the optical response at 1.5 eV that is sensitive to the magnetic order parameter is polarized in the direction normal to the Ir hexagons}. While this clearly can account for the lack of a step in $R$ in NIO at its $T_c$ of 14 K, below we show that same assumption can acccount for the observed anistropy in LIO as well.

To model the component of $R$ in LIO that depends on $S$ we add contributions to the polarizability from the two inequivalent hexagons. We assume that the response of each hexagon can be parameterized by $\alpha_\|$ and $\alpha_\perp$, polarizabilities for $\mathbf{E}$ parallel and perpendicular to the hexagon plane, respectively.  The optical response in the \textit{a-b} plane is the sum of contributions from two sets of hexagons counter-rotated by $\theta = 35^{\circ}$, corresponding to the $70^{\circ}$ angle observed between honeycomb chains in LIO.~\cite{ModicNatCom14} With this assumption we can write the components of the polarizability tensor for directions \textbf{a} and \textbf{b} in terms of $\alpha_\|$ and $\alpha_\perp$,
\begin{equation}
\begin{aligned}
	\alpha_{a} &= \cos^{2}({\theta}) \alpha_{\perp} + \sin^{2}({\theta}) \alpha_{\parallel}\\
	\alpha_{b} &= \sin^{2}({\theta}) \alpha_{\perp} + \cos^{2}({\theta}) \alpha_{\parallel}.
\end{aligned}
\end{equation}
The model predicts a polarizability ratio, $\alpha_a/\alpha_b=\cot^2(35^{\circ})=2.04$, which is consistent with both the direction and magnitude of the observed anistropy. We note that this anisotropy is stronger than is observed in the equilibrium optical conductivity shown in Fig. 2, suggesting that only the component of the optical response that is sensitive to the magnetic order is strongly polarized with respect to the plane of the Ir hexagons.

We conclude this section with a discussion of the other striking feature of the $\Delta R$ anisotropy in LIO, which is the onset of a slow, negative component of $\Delta R_b$ well above the magnetic transition temperature.  This component is highlighted by the shaded regions in the depiction of the 40, 50, and 60 K data in Fig. 8a. The observation that $\Delta R_b (t,T)$ is special for $T>T_c$ is consistent with a unique structural feature of the LIO harmonic honeycomb.  As depicted in Fig. 9, there are two \textit{c}-oriented links in LIO: one that bridges two chains of hexagons and another that forms a bond in each Ir hexagon. The hopping between linked Ir atoms is mediated by the two nearest neighbor O atoms, which all together form a coplanar Ir-O$_2$-Ir unit.  In the $\gamma-$Li$_2$IrO$_3$ structure, as opposed to the layered honeycomb, the normal to this plane is parallel to the $b$ principal axis. By contrast, the IrO$_2$ planes in the layered honeycomb structure are neither parallel nor perpendicular to the layer plane.

\begin{figure}[ht]
	\begin{center}
		\includegraphics[width=8.5cm]{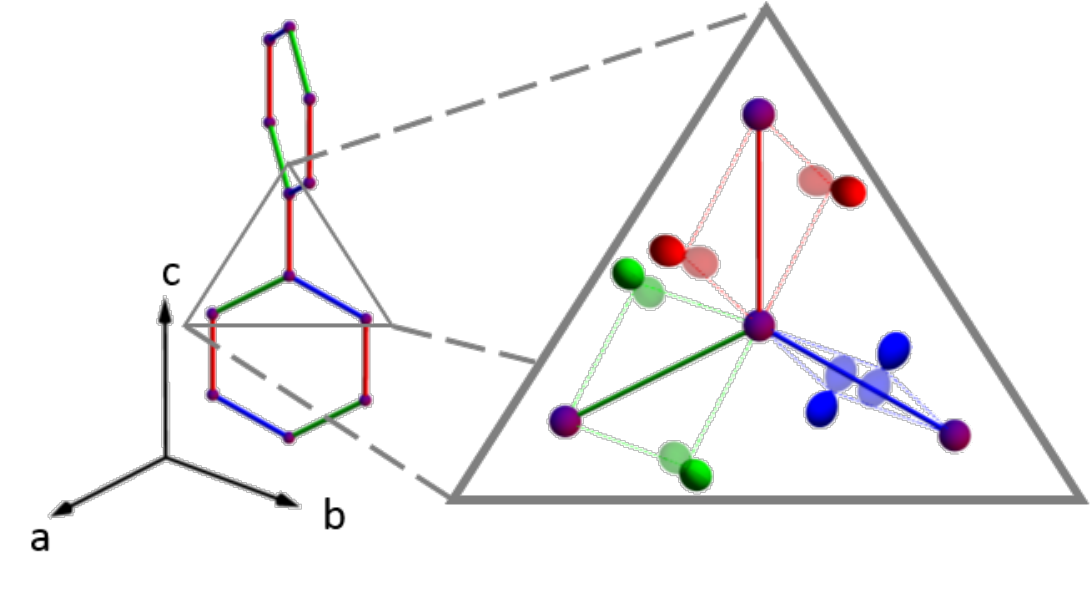}
	\end{center}
	\caption{\label{fig:8} An illustration of the arrangement of two iridium hexagons in LIO that belong to nearest neighbor chains. The $x$, $y$, and $z$ Kitaev bonds are shown in blue, green, and red, respectively, with the principal crystallographic axes in black. The triangular popout shows the oxygen $2p$ orbitals involved in hopping between the neighboring iridium atoms. }
\end{figure}

The fact that in LIO, $\Delta R_b$ deviates from the high-$T$ isotropic signal well above $T_c$ suggests that magnetic correlations develop in the normal state initially on the \textit{b}-axis oriented Ir-O$_2$-Ir planes. This observation is consistent with the magnetic susceptibility, $\chi$, of LIO, where it is found that $\chi_b$ grows much more rapidly than $\chi_a$ and $\chi_c$ as the temperature is lowered towards $T_c$.~\cite{ModicNatCom14}  The nearly divergent $\chi_b$ implies correlations of moments on the \textit{c}-oriented Ir$_2$ links that favor ferromagnetic alignment in the \textit{b}-axis direction.  This is precisely the correlation that is expected to develop from the Kitaev interaction.  When this coupling is sufficiently dominant, the ground state is expected to be a spin liquid, rather than a magnetically ordered state.~\cite{KimchiArxiv14} The transition to magnetic order in LIO indicates that subdominant magnetic interactions destabilize the spin liquid state, leading to the complex spiral magnetism that is seen by X-ray diffraction. The abrupt appearance of a positive component in both $\Delta R_a$ and $\Delta R_b$ at $T_c$ is consistent with the idea that the spin order that onsets at $T_c$ is distinct from the Kitaev-like spin alignment that appears in the normal state.

\section{Summary and Conclusions}

The work reported here is the first study of the equilibrium and transient optical reflectance of the recently synthesized "harmonic honeycomb" or $\gamma$ polytype of Li$_2$IrO$_3$.~\cite{ModicNatCom14} Throughout we have the compared the properties of LIO, in which the basic building block of Ir hexagons form a three dimensional network, with those of NIO, where the hexagons form a layered honeycomb lattice. The equilibrium optical conductivity of LIO indicates that it is an insulator with an optical gap in the range from 0.3-0.5 eV, as expected from measurements of resistivity vs. temperature.~\cite{ModicNatCom14} At photon energies below 1 eV the  optical conductivity of LIO is equal to that of NIO within experimental uncertainty, indicating that the low-energy electronic excitations are very similar, despite their structural differences. Detailed measurements of the transient reflectance of LIO as a function of delay time and temperature reveal a three component picture, with significant anisotropy with respect to the direction of the probe electric field in the \textit{ab} plane. In contrast, the transient reflectance of NIO is isotropic, as expected for reflection from a planar hexagonal structure, and has only one temperature dependent component.

Of the three components observed in LIO, one clearly has the same origin as the single component observed in NIO, as judged from the similiarity of the time delay and temperature dependence. The excited state that gives rise to this component can be characterized as metastable, as once it appears on a 10 ps time scale it survives for several hundred picoseconds. The same photoexcitation state has been seen previously in a transient grating study of NIO,~\cite{AlpischevPRL15} where it was proposed that $\Delta R(t,T)$ at low $T$ arises from low-lying holon-doublon states that are bound by the energy cost of deforming NIO's "zig-zag" magnetic structure. However, given that this component of $\Delta R(t,T)$ is essentially identical in the two compounds despite the fact that LIO manifests incommensurate spiral rather than commensurate ziz-zag order, it is clear that the nature of photoexcited quasiparticles in these hexagonal iridates is not determined by the form of long-range magnetic order. It is likely that the properties of these states derive from the local electronic structure of the edge-sharing Ir-O octahedra. Whether these excitations are topologically trivial, for example spin polarons, or more exotic topologically protected quasiparticles, remains a subject for future research. In particular, measurements of the conductivity of the photoexcited state, either by direct transport or contactless terahertz methods, can determine whether these excitations are charged or neutral. Certainly, understanding the nature of the low-lying photoexcited states takes on a heightened significance if future work confirms that iridates exhibit high-$T_c$ superconductivity when doped.

The components of $\Delta R$ that are unique to LIO are associated with magnetic order. One component onsets discontinuously at $T_c$, with a divergent rise time, indicating that is associated with critical slowing down at a continuous transition to a long-range ordered state. The step-like increase in $\Delta R$ is roughly twice larger for probe electric field parallel to the \textbf{a} than to \textbf{b}.  We showed that this anisotropy is consistent with the hypothesis that the optical transitions in LIO that are sensitive to long-range magnetic order are polarized in the direction perpendicular to the plane of the Ir hexagons.  This same hypothesis can then account for the absence of a critical component of $\Delta R$ in NIO at its critical temperature of 14 K, as in the layered honeycomb structure the probe electric field is oriented parallel to the Ir hexagons. The remaining component of $\Delta R$ in LIO appears only for probe field in the \textbf{b} direction and is observable approximately 15 K above $T_c$. We suggested that this component arises from local magnetic correlations on the \textbf{c}-oriented bonds that link the two sets of Ir hexagonal chains, and is related to the nearly divergent magnetic susceptibility seen above $T_c$ for magnetic fields in the \textbf{b} direction.\cite{ModicNatCom14}

We would like to thank R. Valenti for discussions, as well as H. Bechtel and M. Martin for support at the Advanced Light Source beamline 1.4.3 and 1.4.4. This work was supported by the Director, Office of Science, Office of Basic Energy Sciences, Materials Sciences and Engineering Division, of the U.S. Department of Energy under Contract No. DE-AC02-05CH11231.


\begin{thebibliography}{99}
\bibitem{KimPRL08}
B.~J.~Kim \textit{et al}., Phys.\ Rev.\ Lett.\ {\bf 101}, 076402 (2008).

\bibitem{KimScience09}
B.~J.~Kim \textit{et al}., Science {\bf 323}, 345 (2009).

\bibitem{SenthilPRL}
F.~Wang, and T.~Senthil, Phys.\ Rev.\ Lett.\ {\bf 106}, 136402 (2010).

\bibitem{KimScience14}
Y.~K.~Kim \textit{et al}., Science {\bf 334}, 187 (2014).

\bibitem{Jan15}
Y.~J.~Jan \textit{et al}., arXiv:1506.06557(2015).

\bibitem{Kim15}
Y.~K.~Kim \textit{et al}., arXiv:1506.06639 (2015).

\bibitem{AritaPRL12}
R.~Arita, J.~Kunes, A.~V.~Kozhevnikov, A.~G.~Eguiluz, and M.~Imada, Phys.\ Rev.\ Lett.\ {\bf 108}, 086403 (2012).

\bibitem{CominPRL12}
R.~Comin \textit{et al}., Phys.\ Rev.\ Lett.\ {\bf 109}, 266406 (2012).

\bibitem{HaskelPRL12}
D.~Haskel \textit{et al}., Phys.\ Rev.\ Lett.\ {\bf 109}, 027204 (2012).

\bibitem{HseihPRB12}
D.~Hsieh, F.~Mahmood, D.~H.~Torchinsky, G.~Cao, and N.~Gedik, Phys.\ Rev.\ B\ {\bf 86}, 035128 (2012).

\bibitem{MazinPRL12}
I.~I.~Mazin, H.~O.~Jeschke, K.~Foyevtsova, R.~Valenti, and D.~I.~Khomskii, Phys.\ Rev.\ Lett.\ {\bf 109}, 197201 (2012).

\bibitem{MazinPRB13}
I.~I.~Mazin \textit{et al}., Phys.\ Rev.\ B\ {\bf 88}, 035115 (2013).

\bibitem{SohnPRB13}
C.~H.~Sohn \textit{et al}., Phys.\ Rev.\ B\ {\bf 88}, 085125 (2013).

\bibitem{ChaloupkaPRL10}
J.~Chaloupka, G.~Jackeli, and G.~Khaliullin, Phys.\ Rev.\ Lett.\ {\bf 105}, 027204 (2010).

\bibitem{SinghPRL12}
Y.~Singh \textit{et al}., Phys.\ Rev.\ Lett.\ {\bf 108}, 127203 (2012).

\bibitem{GretarssonPRL13}
H.~Gretarsson \textit{et al}., Phys.\ Rev.\ Lett.\ {\bf 110}, 076402 (2013).

\bibitem{LiuPRB11}
X.~Liu \textit{et al}., Phys.\ Rev.\ B\ {\bf 83}, 220403(R) (2011).

\bibitem{ChoiPRL12}
S.~K.~Choi \textit{et al}., Phys.\ Rev.\ Lett.\ {\bf 108}, 127204 (2012).

\bibitem{ChunNatPhys15}
S.~H.~Chun \textit{et al}., Nature Phys.\ {\bf 11}, 462 (2015).

\bibitem{YePRB12}
F.~Ye \textit{et al}., Phys.\ Rev.\ B\ {\bf 85}, 180403(R) (2012).

\bibitem{FoyevtsovaPRB13}
K.~Foyevtsova, H.~O.~Jeschke, I.~I.~Mazin, D.~I.~Khomskii, and R.~Valenti, Phys.\ Rev.\ B\ {\bf 88}, 035107 (2013).

\bibitem{KimSciRep14}
H.-J.~Kim, J.-H.~Lee and J.-H.~Cho, Sci.\ Rep.\ {\bf 4}, 5253 (2014).

\bibitem{TakayamaArxiv14}
T.~Takayama \textit{et al}., arXiv:1403.3296 (2014).

\bibitem{ModicNatCom14}
K.~A.~Modic \textit{et al}., Nat.\ Commun.\ {\bf 5}, 4203 (2014).

\bibitem{BiffinPRL14}
A.~Biffin \textit{et al}., Phys.\ Rev.\ Lett.\ {\bf 113}, 197201 (2014).

\bibitem{KimchiPRB14}
I.~Kimchi, J.~G.~Analytis and A.~Vishwanath, Phys.\ Rev.\ B\ {\bf 90}, 205126 (2014).

\bibitem{KimchiArxiv14}
I.~Kimchi, R.~Coldea and A.~Vishwanath, arXiv:1408.3640v2 (2014).

\bibitem{LeeArxiv14}
E.~K.-H.~Lee and Y.~B.~Kim, arXiv:1407.4125v3 (2014).

\bibitem{ReutherArxiv14}
J.~Reuther, R.~Thomale and S.~Rachel, arXiv:1404.5818v2 (2014).

\bibitem{SinghPRB10}
Y.~Singh and P.~Gegenwart, Phys.\ Rev.\ B\ {\bf 82}, 064412 (2010).

\bibitem{LiArxiv14}
Y.~Li \textit{et al}., arXiv:1410.4243 (2014).

\bibitem{BeaurepairePRL96}
E.~Beaurepaire, J.~C.~Merle, A.~Daunois, and J.~Y.~Bigot, Phys.\ Rev.\ Lett.\ {\bf 76}, 4250 (1996).

\bibitem{KoopmansNatMat10}
B.~Koopmans \textit{et al}., Nature Mater.\ {\bf 9}, 259 (2009).

\bibitem{KisePRL00}
T.~Kise \textit{et al}., Phys.\ Rev.\ Lett.\ {\bf 85}, 1986 (2000).

\bibitem{OgasawaraPRL05}
T.~Ogasawara \textit{et al}., Phys.\ Rev.\ Lett.\ {\bf 94}, 087202 (2005).

\bibitem{KantnerPRB11}
C.~L.~S.~Kantner \textit{et al}., Phys.\ Rev.\ B\ {\bf 83}, 134432 (2011).

\bibitem{KoopmansPRL05}
B.~Koopmans, J.~J.~M.~Ruigrok, F.~Dalla~Longa, and W.~J.~M.~de~Jonge, Phys.\ Rev.\ Lett.\ {\bf 95}, 267207 (2005).

\bibitem{AlpischevPRL15}
Z.~Alpichshev, F.~Mahmood, G.~Cao, and N.~Gedik, Phys.\ Rev.\ Lett.\ {\bf 114}, 017203 (2015).

\end{thebibliography}
\end{document}